\documentclass[showpacs,preprint,amsmath,amssymb]{revtex4}
\usepackage{graphicx}
\usepackage{dcolumn}
\usepackage{epsfig}
\usepackage{bm}

\begin{document}
\title{The Paschos-Wolfenstein relation in a hadronic picture}
\author{C.~Praet}
\email{christophe.praet@UGent.be}
\author{N.~Jachowicz} 
\email{natalie.jachowicz@UGent.be}
\author{J.~Ryckebusch}
\author{P.~Vancraeyveld}
\author{K.~Vantournhout}
\affiliation{Department of Subatomic and Radiation Physics,\\ Ghent University, \\Proeftuinstraat 86, \\ B-9000 Gent, Belgium.}
\date{\today}
\begin{abstract}
The Paschos-Wolfenstein (PW) relation joins neutral- and
charged-current neutrino- and antineutrino-induced cross sections into an expression that
depends on the weak mixing angle $\sin^2 \theta_W$.  Contrary to the traditional approach with
partonic degrees of freedom, we adopt a model built on
hadronic degrees of freedom to perform a study of the PW relation at
intermediate neutrino energies ($100$\ MeV - $2$\ GeV).  With upcoming
high-statistics scattering experiments like MINER$\nu$A and FINeSSE, a scrutiny of the PW relation is timely.  Employing a
relativistic Glauber nucleon knockout model for the
description of quasi-elastic neutrino-nucleus reactions, the influence
of nuclear effects on the PW relation is investigated.  We discuss nuclear model
dependences and show that the PW relation is a robust ratio,
mitigating the effect of final-state interactions, for example, to the
1\% level.  The role played by a possible strangeness
content of the nucleon is investigated.  It appears that the
uncertainties arising from the poorly known strangeness parameters
and the difficulties in nuclear modelling seriously limit the applicability 
of the PW relation as an intermediate-energy electroweak precision
tool.  On the other hand, we show that nuclear effects may be
sufficiently well under control to allow the extraction of new
information on the axial strangeness parameter.  Results are presented for $^{16}\mbox{O}$ and $^{56}\mbox{Fe}$. 
\end{abstract}

\maketitle

\section{Introduction}
Now more than ever, neutrinos are valued for their wide probing
potential in many different domains.  At intermediate energies, they
are put forward to study nucleon structure and probe nuclear effects
\cite{Minerva1}.  Well-defined ratios of neutrino-scattering
cross sections prove being promising tools to measure the strange-quark
contribution to the nucleon spin \cite{Pate1, Finesse1}.  Lately,
neutrinos have been regarded as interesting candidates for electroweak
tests aiming at a precision measurement of the Weinberg angle
$\theta_W$ \cite{Zeller1, Balantekin1, Conrad1}.\\
One of the most fundamental parameters in the standard model (SM), the
weak mixing angle has been at the center of research activities, involving
both theoretical SM calculations \cite{Czarnecki1, Musolf1} and experimental
efforts to determine its value.  While all $\sin^2 \theta_W$ measurements 
near the $Z^0$ pole \cite{LEP1, SLD1} and for low $Q^2$ values
\cite{APV1, Anthony1} are in good agreement with the SM prediction, 
an experiment by the NuTeV collaboration at $Q^2 = 20\
\mbox{GeV}^2$ does not seem to corroborate the calculated 
running of the Weinberg angle \cite{Zeller1}.  Explanations for this 
anomalous result range from quantum chromodynamics (QCD) uncertainties 
\cite{Zeller2, Diener1}, to nuclear effects \cite{Eskola1, Kulagin1}
and even interpretations involving new physics \cite{Kurylov1,
  Davidson1}.  Whether the surprising NuTeV outcome can
be resolved through a further analysis of the data or indeed hints at new
physics beyond the SM, is up to this day an unresolved issue
\cite{Mohapatra1}.  In NuTeV's analysis, the Paschos-Wolfenstein
relation \cite{Paschos1} plays an essential role in relating the weak 
mixing angle to measured ratios of neutral-current (NC) to
charged-current (CC) deep-inelastic scattering (DIS) neutrino cross 
sections.  As a consequence, it has been tested very well in the DIS
regime with respect to genuine QCD mechanisms.  To the contrary, little effort 
has been put in the intermediate-energy regime, where an adequate
description in terms of hadronic rather than partonic degrees of
freedom is needed.\\
In this work, we explore what physics could be probed by future
measurements of the Paschos-Wolfenstein relation at medium energies.  
With newly proposed, high-precision neutrino-scattering experiments 
like MINER$\nu$A \cite{Minerva1} and FINeSSE \cite{Finesse1}, it is
timely to make predictions about the level of sensitivity one
would need to extract relevant physics from these measurements.  As a
matter of fact, the MINER$\nu$A proposal contains an extensive program
for studying nuclear effects with neutrinos \cite{Minerva2}.  More
specifically, the impact of the nuclear medium on NC/CC cross-section
ratios will be investigated by employing carbon, iron and lead target
nuclei.  In this paper, we focus on a study of the PW
relation in the few GeV regime, adopting a model based on hadronic
degrees of freedom \cite{Lava1}.  Considering quasi-elastic (QE) 
neutrino-nucleus scattering with nucleon knockout as the basic 
source of strength in the $100$\ MeV - $2$\ GeV energy range, the PW
relation is constructed for both oxygen and iron target nuclei.
Treating nucleon-nucleon interactions in a relativistic mean-field
approximation, binding effects and the Pauli exclusion principle are 
naturally included in our approach.  Final-state interactions 
of the outgoing nucleon are incorporated through a Glauber approach.  
Within this model, we show how the nuclear medium affects the PW
relation.  A model-dependence discussion is included in this work, 
by comparing predictions within different frameworks.\\
Knowing at what level nuclear uncertainties affect the PW relation, one
can proceed with putting theoretical constraints on the accuracy
with which variables can be determined from it.  In earlier work by Donnelly and
Musolf \cite{Donnelly1}, nuclear uncertainties were
estimated too large to allow a $\sin^2 \theta_W$ determination in
parity-violating electron scattering (PVES) with a precision similar to
other types of measurements.  It is important to check if the PW
relation at medium energies provides a powerful tool for a
Weinberg-angle extraction in the QE regime.  In addition, the
Paschos-Wolfenstein relation has been suggested 
to serve as a lever for the determination of the strange-quark
contribution to the nucleon's spin, $g_A^s$ \cite{Alberico1,
  Alberico2}.  This and other work \cite{Alberico3, Barbaro1,
  Maieron1} points out that for sufficiently high energies ($\sim
1$\ GeV), ratios of neutrino cross sections can serve as theoretically
clean probes for the nucleon's strangeness content.  Here, we derive a 
theoretical error bar for $g_A^s$ as extracted from the PW relation.
Given that the PW relation is both sensitive to the weak
mixing angle and the strangeness content of the nucleon, it is
worthwhile to conduct a study of how these parameters are
intertwined.  This type of study is surely relevant for the
future FINeSSE experiment, which aims at measuring the ratios of NC to CC 
neutrino-induced cross sections at medium energies to extract 
information on the strange axial form factor $g_A^s$.\\  
The paper is organized as follows.  Section \ref{pw} introduces the
Paschos-Wolfenstein relation in its traditional, DIS form.  The third
section discusses the theoretical framework used in this paper for the 
description of neutrino-nucleus interactions.  An analytical estimate 
of the Paschos-Wolfenstein ratio for intermediate-energy
neutrino-nucleus scattering reactions is derived in section
\ref{pwnuc}.  Numerical results are presented in section
\ref{results}.  Our conclusions are summarized in section VI.
                  
\section{The Paschos-Wolfenstein relation}\label{pw}
Traditionally, the Paschos-Wolfenstein relation is defined as the
following ratio of NC to CC (anti)neutrino-nucleon cross sections
\begin{equation}
\label{PaWo}
\mbox{PW} = \frac{\sigma^{\mathrm{NC}}(\nu N) - \sigma^{\mathrm{NC}}(\overline{\nu} N)}{\sigma^{\mathrm{CC}}(\nu N) - \sigma^{\mathrm{CC}}(\overline{\nu} N)}.
\end{equation}
Adopting the nucleon's quark-parton structure, the PW relation can be computed starting from the quark currents  
\begin{equation}
\label{pw1}
\begin{split}
\hat{\j}^{(Z)}_{\mu} = \sum_{q=u,d} g_{q,L} \overline{q} \gamma_{\mu} (1-\gamma_5) q + g_{q,R}\overline{q} \gamma_{\mu}(1+\gamma_5)q & \quad \mathrm{NC},\\
\hat{\j}^{(+)}_{\mu} = \frac{1}{2}\overline{u}\gamma_{\mu}(1-\gamma_5)d,\quad \hat{\j}^{(-)}_{\mu} = \frac{1}{2}\overline{d}\gamma_{\mu}(1-\gamma_5)u & \quad \mathrm{CC},
\end{split}
\end{equation}
with the quark coupling strengths
\begin{equation}
\label{pw2}
\begin{split}
g_{u,L} = & \frac{1}{2} - \frac{2}{3} \sin^2 \theta_W,\quad g_{u,R} = -\frac{2}{3} \sin^2 \theta_W,\\
g_{d,L} = - & \frac{1}{2} + \frac{1}{3} \sin^2 \theta_W,\quad g_{d,R} = \frac{1}{3} \sin^2 \theta_W. 
\end{split}
\end{equation}
Using these expressions, one immediately derives \cite{Zeller3}
\begin{equation}
\label{pw5}
\mbox{PW} = \left(\frac{1}{\cos^2\theta_c}\right)\left( \frac{1}{2} - \sin^2\theta_W \right),
\end{equation}       
where $\theta_c$ stands for the Cabibbo mixing angle.  Equation
(\ref{pw5}) holds for isoscalar targets, containing an equal number of
$u$ and $d$ quarks, and neglecting $s$ quarks.

\section{Cross sections for quasi-elastic neutrino-nucleus
  interactions}\label{cs}
A description in terms of quark currents is no longer appropriate when
considering neutrino-nucleus interactions at medium energies.  Instead, one usually
invokes form factors to map the nucleon's substructure.  With these
form factors, matrix elements of the hadronic current are constructed
based on general principles of Lorentz invariance.  In this section, the formalism
employed for the calculation of neutrino-nucleus cross sections is
presented.  We consider quasi-elastic (anti)neutrino-nucleus interactions of the following type
\begin{equation}
\label{process}
\begin{split}
\nu + A & \stackrel{\mbox{NC}}{\longrightarrow} \nu + (A-1) + N,\\
\overline{\nu} + A & \stackrel{\mbox{NC}}{\longrightarrow} \overline{\nu} + (A-1) + N,\\
\nu + A & \stackrel{\mbox{CC}}{\longrightarrow} l^- + (A-1) + p,\\
\overline{\nu} + A & \stackrel{\mbox{CC}}{\longrightarrow} l^+ + (A-1) + n,
\end{split} 
\end{equation}
limiting ourselves to processes where the final nucleus $(A-1)$ is left with an excitation energy not exceeding a few tens of MeV.  The target nucleus is denoted by its mass number $A$, $l$ represents an outgoing charged lepton and $N$ stands for the ejectile (proton $p$ or neutron $n$).  To calculate the corresponding cross sections, we turn to the relativistic quasi-elastic nucleon knockout model described in \cite{Lava1}.  Writing $K'^{\mu} = (\epsilon',\vec{k}')$, $K^{\mu}_N = (\epsilon_N,\vec{k}_N)$ and $K_{A-1}^{\mu} = (\epsilon_{A-1},\vec{k}_{A-1})$ for the four-momenta of the scattered lepton, the ejectile and the residual nucleus, these cross sections are given by      
\begin{equation}
\label{form1}
\frac{d^5\sigma}{d\epsilon'd^2\Omega_ld^2\Omega_N} = \frac{M_lM_NM_{A-1}}{(2\pi)^5\epsilon'}k'^2k_Nf^{-1}_{rec} \overline{\sum_{if}} |M_{fi}|^2.
\end{equation}
The exclusive cross section (\ref{form1}) still depends on the solid angles $\Omega_l$ and $\Omega_N$, determining the direction of the scattered lepton and ejectile respectively.  The hadronic recoil factor $f_{rec}$ is given by
\begin{equation}
\label{form2}
f_{rec} = \left| \epsilon_{A-1} + \epsilon_N(1-\frac{\vec{q} \cdot \vec{k}_N}{k_N^2})\right|.
\end{equation}
Further on, an appropriate averaging over initial states and sum over
final states is performed in the squared invariant matrix element
$|M_{fi}|^2$.  Using the Feynman rules, one finds
\begin{equation}
\label{form3}
|M_{fi}|^2 = \frac{g^4}{64 \frac{M_W^4}{M_{Z,W}^4}(Q^2+M_{Z,W}^2)^2}l_{\alpha\beta}W^{\alpha\beta},
\end{equation}  
with $g$ the weak coupling strength and $Q^2 = - q_{\mu} q^{\mu}$ the
four-momentum transfer.  For NC (CC) interactions, the boson mass
$M_Z$ ($M_W$) is selected.  In the CC case, the right-hand side of
(\ref{form3}) should also be multiplied by $\cos^2\theta_c$.  One further distinguishes a lepton part described by the tensor $l_{\alpha\beta}$ and a nuclear part, described by the tensor
\begin{equation}
\label{form4}
W^{\alpha\beta} = (\mathcal{J}^{\alpha})^{\dagger} \mathcal{J}^{\beta}.
\end{equation}
To evaluate the nuclear current matrix elements $\mathcal{J}^{\mu}$,
we assume that the major fraction of the transferred energy is carried
by the ejectile, thereby neglecting processes that involve
several target nucleons.  In the impulse approximation (IA), 
the nuclear many-body current operator is replaced by a sum of one-body current operators $\hat{J}^{\mu}$
\begin{equation}
\sum_{k=1}^{A} \hat{J}^{\mu}(\vec{r}_k).
\end{equation}
Employing an independent-particle model (IPM) for the initial and final nuclear wave functions, the current matrix elements can be written as \cite{Lava1}
\begin{equation}
\mathcal{J}^{\mu} = \int d \vec{r} \overline\phi_F (\vec{r}) \hat{J}^{\mu}(\vec{r}) e^{i \vec{q} \cdot \vec{r}} \phi_B (\vec{r}),
\end{equation}
where $\phi_B$ and $\phi_F$ are relativistic bound-state and scattering wave functions.  We adopt the following expression for the weak one-nucleon current operator
\begin{equation}
\label{form5}
\begin{split}
\hat{J}^{\mu} =\ & F_1(Q^2) \gamma^{\mu} +
  \frac{i}{2M_N} F_2(Q^2)\sigma^{\mu\nu}q_{\nu} \\
  +\ & G_A(Q^2) \gamma^{\mu} \gamma_5 + \frac{1}{2M_N} G_P(Q^2)
  q^{\mu} \gamma_5,
\end{split}
\end{equation} 
composed of a vector part, described by the Dirac and Pauli form
factors $F_1$ and $F_2$, and an axial part, described by the axial
form factor $G_A$ and pseudoscalar form factor $G_P$.  As pointed out
for example in \cite{Lava1}, one can choose amongst different options for the one-body 
vertex function, of which (\ref{form5}) is labeled \textit{cc2}.  For
bound nucleons these parameterizations do not produce identical results, giving rise to the
so-called Gordon ambiguity.  For the vector form factors two different
parameterizations will be considered: a standard dipole form and the
BBA parameterization of Ref.~\cite{Budd1}.  The axial form factor
$G_A$ will be parameterized by a dipole.  Using the Goldberger-Treiman
relation, the pseudoscalar form factor can be related to the axial one  
\begin{equation}
G_P(Q^2) = \frac{2 M_N}{Q^2 + m_{\pi}^2} G_A(Q^2),
\end{equation}
with $m_{\pi}$ the pion mass.  As the contribution of $G_P$ to the cross section is proportional to the scattered lepton's mass, it vanishes for NC reactions.  At $Q^2=0$, the form-factor values are given by 
\begin{equation}
\label{form6}
G_A = \begin{cases} \frac{-g_A\tau_3 + g_A^s}{2} & \mbox{NC} \\
g_A \tau_{\pm} & \mbox{CC} 
\end{cases}
\end{equation}
and
\begin{equation}
\label{form6bis}
F_i = \begin{cases} (\frac{1}{2}-\sin^2\theta_W)F^{EM,V}_i\tau_3 & \\
 - \sin^2\theta_W F^{EM,S}_i - \frac{1}{2} F^s_i &\ \ \ \mbox{NC}\\
 & \\
F^{EM,V}_i \tau_{\pm} &\ \ \ \mbox{CC},
\end{cases}
\end{equation}
where the superscript $s$ refers to strangeness contributions, $g_A = 1.262$ and the isospin operators are defined in the standard way as
\begin{equation}
\begin{split}
\tau_3 \vert p \rangle & = + \vert p \rangle,\quad \tau_3 \vert n \rangle = - \vert n \rangle,\\
\tau_+ \vert n \rangle & = + \vert p \rangle,\quad \tau_+ \vert p \rangle = 0,\\
\tau_- \vert p \rangle & = - \vert n \rangle,\quad \tau_- \vert n \rangle = 0.
\end{split}
\end{equation}
The relation between the weak vector form factors and the
electromagnetic isovector $F_i^{EM,V} = F_{i,p}^{EM} - F_{i,n}^{EM}$
and isoscalar $F_i^{EM,S} = F_{i,p}^{EM} + F_{i,n}^{EM}$ ones is established by the conserved vector-current (CVC) hypothesis.\\
Combining terms into longitudinal, transverse and interference contributions, the cross section for NC interactions in Eq. (\ref{form1}) can be written as
\begin{equation}
\label{form12}
\begin{split}
\frac{d^5\sigma}{d\epsilon'd^2\Omega_l d^2\Omega_N} = & \frac{M_N M_{A-1}}{(2\pi)^3} k_N f_{rec}^{-1} \sigma^Z \left[ v_L R_L + v_T R_T \right.\\
& + v_{TT} R_{TT} \cos2\phi + v_{TL} R_{TL} \cos\phi \\ 
& \left. \pm (v'_T R'_T + v'_{TL} R'_{TL} \cos\phi)\right],
\end{split}
\end{equation} 
where the upper (lower) sign relates to antineutrino (neutrino) cross sections.  We use the notation
\begin{equation}
\sigma^Z = \left(\frac{G_F \cos(\theta_{l}/2)\epsilon'M_Z^2}{\sqrt{2}\pi(Q^2+M_Z^2)} \right)^2,
\end{equation}
and the definitions of Table~I.  The lepton scattering angle is denoted by $\theta_l$, whereas $\phi$ stands for the azimuthal angle between the lepton scattering plane and the hadronic reaction plane, defined by $\vec{k}_N$ and $\vec{q}$.
\begin{table*}
\caption[table1]{Kinematic factors and response functions for NC and
  CC (anti)neutrino-nucleus scattering.  Hadronic matrix elements are
  expressed in the spherical basis $\vec{e}_z$, $\vec{e}_{\pm1} = \mp
  \frac{1}{\sqrt{2}} (\vec{e}_x \pm i \vec{e}_y)$, $\mathcal{J}^{\mu}
  = (\mathcal{J}^0,\vec{\mathcal{J}})$ with $\vec{\mathcal{J}} =
  -\mathcal{J}^{-1} \vec{e}_{+1} - \mathcal{J}^{+1} \vec{e}_{-1} +
  \mathcal{J}^z \vec{e}_z$.  For the CC case, we only list those
  expressions that differ from the NC ones.}
\ \\[1mm]
\begin{tabular}{rcl|rcl}\hline
\multicolumn{3}{c}{\rule[-2mm]{0mm}{6mm}\textbf{Kinematic factors}} &
\multicolumn{3}{c}{\textbf{Response functions}}\\ \hline
\multicolumn{6}{l}{\rule[-1mm]{0mm}{4mm} \textit{Neutral current}}\\ \hline
$v_L$ & = & $1$ & $R_L$ & = & $\left| \mathcal{J}^0 - \frac{\omega}{q} \mathcal{J}^z \right|^2$\\ 
$v_T$ & = & $\tan^2 \frac{\theta_{l}}{2} + \frac{Q^2}{2q^2}$ & $R_T$ & = & $\left| \mathcal{J}^{+1} \right|^2 + \left| \mathcal{J}^{-1} \right|^2$\\ 
$v_{TT}$ & = & $- \frac{Q^2}{2q^2}$ & $R_{TT} \cos 2\phi$ & = & $2 \Re \left((\mathcal{J}^{+1})^{\dagger} \mathcal{J}^{-1}\right)$ \\
$v_{TL}$ & = & $- \frac{1}{\sqrt{2}}\sqrt{\tan^2 \frac{\theta_{l}}{2} + \frac{Q^2}{q^2}}$\hspace{1cm} & $R_{TL}\cos\phi$ & = & $-2\Re\left(\mathcal{J}^0 - \frac{\omega}{q} \mathcal{J}^z \right)(\mathcal{J}^{+1} - \mathcal{J}^{-1})^{\dagger}$\\ 
$v'_T$ & = & $\tan\frac{\theta_l}{2}\sqrt{\tan^2 \frac{\theta_{l}}{2} + \frac{Q^2}{q^2}}$ & $R'_T$ & = & $\left| \mathcal{J}^{+1} \right|^2 - \left| \mathcal{J}^{-1} \right|^2$\\
\rule[-3mm]{0mm}{6mm} $v'_{TL}$ & = & $\frac{1}{\sqrt{2}}\tan\frac{\theta_{l}}{2}$ & $R'_{TL}\cos\phi$ & = & $-2\Re\left(\mathcal{J}^0 - \frac{\omega}{q} \mathcal{J}^z \right)(\mathcal{J}^{+1} + \mathcal{J}^{-1})^{\dagger}$\\ \hline
\multicolumn{6}{l}{\rule[-1mm]{0mm}{4mm} \textit{Charged current}}\\ \hline
$v_L R_L$ & = & \multicolumn{4}{l}{$\left(1 + \zeta \cos\theta_{l}\right)|\mathcal{J}^0|^2 + \left(1 + \zeta \cos\theta_{l} - \frac{2\epsilon\epsilon'}{q^2} \zeta^2 \sin^2\theta_{l}\right)|\mathcal{J}^z|^2$}\\
& & \multicolumn{4}{l}{$-\left(\frac{\omega}{q} \left(1+ \zeta \cos\theta_{l} \right) + \frac{M_l^2}{\epsilon' q} \right) 2 \Re (\mathcal{J}^0 (\mathcal{J}^{z})^{\dagger})$}\\
$v_T$ & = & \multicolumn{4}{l}{$1 -  \zeta \cos\theta_{l} + \frac{\epsilon\epsilon'}{q^2} \zeta^2 \sin^2\theta_{l}$} \\
$v_{TT}$ & = & \multicolumn{4}{l}{$-\frac{\epsilon\epsilon'}{q^2} \zeta^2 \sin^2\theta_{l}$}\\
$v_{TL}R_{TL}\cos\phi$ & = & \multicolumn{4}{l}{$\frac{\sin \theta_{l}}{\sqrt{2}q} (\epsilon + \epsilon') \left( 2 \Re \left( \left( \mathcal{J}^0 - \frac{\omega}{q} \mathcal{J}^z\right)(\mathcal{J}^{+1}  - \mathcal{J}^{-1})^{\dagger} - \frac{M_l^2}{q} \mathcal{J}^z (\mathcal{J}^{+1} - \mathcal{J}^{-1})^{\dagger}\right) \right)$}\\ 
$v'_T$ & = & \multicolumn{4}{l}{$\frac{\epsilon + \epsilon'}{q} \left( 1 -  \zeta \cos\theta_{l} \right) - \frac{M_l^2}{\epsilon' q}$}\\
$v'_{TL}$ & = & \multicolumn{4}{l}{\rule[-3mm]{0mm}{6mm}$-\frac{\sin\theta_{l}}{\sqrt{2}}  \zeta $}\\ \hline
\end{tabular}
\end{table*} 
Due to the non-vanishing mass of the outgoing lepton, CC processes imply expressions that are slightly more involved.  The expressions for the kinematic factors and response functions are listed in the lower part of Table~I.  Furthermore, $\sigma^Z$ has to be replaced by $\sigma^{W\pm}$ where
\begin{equation}
\sigma^{W^{\pm}} = \left(\frac{G_F \cos(\theta_c)\epsilon'M_W^2}{2\pi(Q^2+M_W^2)} \right)^2 \zeta,\quad \zeta = \sqrt{1-\frac{M_l^2}{\epsilon'^2}}.
\end{equation}
Final-state interactions (FSI) of the ejectile with the residual nucleus are taken into account by means of a relativistic multiple-scattering Glauber approximation (RMSGA).  In this approach, the scattering wave function of the outgoing nucleon takes on the form
\begin{equation}
\label{eq:Gl1}
\phi_F(\vec{r}) = G(\vec{b},z)\ \phi_{k_N,s_N}(\vec{r}),
\end{equation}
where $\phi_{k_N,s_N}$ is a relativistic plane wave and $G(\vec{b},z)$
represents the scalar Dirac-Glauber phase.  As a
multiple-scattering extension of the eikonal approximation, the
Glauber approach describes the emission of a fast nucleon from a
composite system of $\mbox{A}-1$ temporarily frozen nucleons.  Details
about the RMSGA approach can be found in Ref.~\cite{Ryckebusch1}.  When FSI are neglected,
$G(\vec{b},z)$ is put equal to $1$, which corresponds to the
relativistic plane-wave impulse approximation (RPWIA).  

\section{Paschos-Wolfenstein relation in neutrino-nucleus scattering}\label{pwnuc}

The cross sections in Eq.~(\ref{form12}) constitute the ingredients
  for our study of the PW relation with hadronic degrees of freedom:
\begin{equation}
\mbox{PW} = \frac{\sigma^{NC}(\nu A) - \sigma^{NC}(\overline{\nu}
  A)}{\sigma^{CC}(\nu A) - \sigma^{CC}(\overline{\nu} A)}. 
\end{equation}
  A numerical calculation of the PW relation can now be performed to
  investigate its behavior with respect to Eq.~(\ref{pw5}) and show
  its sensitivity to various nuclear effects in the
  intermediate-energy range.  Before doing so, however, it is
  interesting to investigate whether the $\sin^2 \theta_W$ dependence of
  Eq.~(\ref{pw5}) can be retrieved within a hadronic picture.  First, for inclusive
  neutrino-scattering reactions, an integration over all angles
  $\Omega_l$, $\Omega_N$ is performed in Eq.~(\ref{form12}), thereby
  nullifying all $\phi$-dependent terms.  Moreover, ignoring the small
  differences between proton and neutron wave functions when
  evaluating the difference of $\nu$- and $\overline{\nu}$-induced
  cross sections, we retain only the contribution from the transverse
  $R'_T$ response.  Obviously, for NC processes, this contribution has
  to be considered for protons and neutrons separately, whereas in the
  denominator, the charge-exchange feature of the interaction forces neutrinos to interact with neutrons and antineutrinos with protons.  Expressing the differential cross sections in terms of the outgoing nucleon's kinetic energy $T_N$, we obtain for an isoscalar nucleus
\begin{widetext}
\begin{equation}
\label{vier3}
\begin{split}
& \frac{\frac{d\sigma^{NC}(\nu
      A)}{dT_N}-\frac{d\sigma^{NC}(\overline{\nu}
      A)}{dT_N}}{\frac{d\sigma^{CC}(\nu
      A)}{dT_N}-\frac{d\sigma^{CC}(\overline{\nu} A)}{dT_N}} \approx \left(\frac{1}{\cos^2\theta_c}\right) \\
& \times \frac{\sum_{\tau_3=\pm1}\sum_{\alpha}\int_0^{\pi} \sin\theta_{l}\sin^2\frac{\theta_{l}}{2} d\theta_{l}\int_0^{\pi} \sin\theta_{N}d\theta_{N}\ k_N\ f_{rec}^{-1}\ \frac{dT_N}{d\epsilon'}\ \frac{\epsilon'^2 M_Z^4}{(4\epsilon\epsilon'\sin^2\frac{\theta_{l}}{2} +M_Z^2)^2}\ \frac{\epsilon+\epsilon'}{q}(R'_T)^{NC}}{\sum_{\alpha}\int_0^{\pi} \sin\theta_{l} \sin^2\frac{\theta_{l}}{2} d\theta_{l}\int_0^{\pi} \sin\theta_{N}d\theta_{N}\ k_N\ f_{rec}^{-1}\ \frac{dT_N}{d\epsilon'}\ \frac{\epsilon'^2M_W^4}{(4\epsilon\epsilon'\sin^2\frac{\theta_{l}}{2}+M_W^2)^2}\ \frac{\epsilon+\epsilon'}{q} (R'_T)^{CC}},
\end{split}
\end{equation} 
\end{widetext}
where the summation over $\alpha$ extends over all bound proton
single-particle levels in the target nucleus.  Furthermore, the mass
of the outgoing lepton has been neglected in Eq.~(\ref{vier3}).
Clearly, the main difference between numerator and denominator lies in
the value of the remaining transverse response function $R'_T$,
proportional to $G_A(Q^2)G_M(Q^2)$ with $G_M = F_1 + F_2$ the magnetic
Sachs form factor.  Assuming that $Q^2 \ll M_Z^2, M_W^2$ and
disregarding differences in the contributions of different shells, the expressions in numerator and denominator cancel to a large extent.  In other words, the PW relation is approximately given by
\begin{widetext}
\begin{equation}
\label{vier6}
\begin{split}
\frac{\frac{d\sigma^{NC}(\nu A)}{dT_N}-\frac{d\sigma^{NC}(\overline{\nu} A)}{dT_N}}{\frac{d\sigma^{CC}(\nu A)}{dT_N}-\frac{d\sigma^{CC}(\overline{\nu} A)}{dT_N}}& \approx \left( \frac{1}{\cos^2\theta_c}\right) \frac{\sum_{\tau_3=\pm1}G_A^{NC}(0)G_M^{NC}(0)}{G_A^{CC}(0)G_M^{CC}(0)}\\
&= \left(\frac{1}{\cos^2\theta_c}\right)\left( (\frac{1}{2} - \sin^2\theta_W) + \frac{g_A^s}{g_A} \left( \frac{\sin^2\theta_W(\mu_p+\mu_n) + \frac{1}{2} \mu_s}{(\mu_p - \mu_n)} \right)\right).
\end{split}
\end{equation}
\end{widetext}
Apart from the standard value figuring in Eq.~(\ref{pw5}), an additional
strangeness term appears.  In (\ref{vier6}), $\mu_p = F_{2,p}^{EM}(0)$
($\mu_n = F_{2,n}^{EM}(0)$) denotes the proton (neutron) magnetic moment
and $\mu_s = F_2^s(0)$ is the strangeness
magnetic moment.  We wish to stress that the left-hand side of
Eq.~(\ref{vier6}) is $T_N$ independent.  

\section{Results and discussion}\label{results}
In the previous section, the DIS expression of the PW relation was
regained by making various approximations to our hadronic picture.  
Next, we will evaluate numerically to what extent the nuclear medium 
affects this \textit{standard value} of the PW relation.  To this end,
the previously neglected nuclear effects are gradually included and
the resulting PW curves are compared with the expression (\ref{vier6}).  First, the strangeness content of the nucleon will be
ignored, putting $g_A^s = 0$ and $\mu_s = 0$.  A discussion of the 
strangeness sensitivity of the PW relation is postponed to subsection~\ref{Strange}.  Results will be presented for $\nu_e$ ($\overline{\nu}_e$)
scattering off both an isoscalar nucleus,
$^{16}_{\phantom{0}8}\mbox{O}$, and a heavier one,
$^{56}_{26}\mbox{Fe}$, with neutron excess.  As a starting
point, we use dipole vector and axial form factors, the \textit{cc2}
form for the one-nucleon current and an on-shell weak mixing angle $\sin^2 \theta_W = 0.2224$.
\subsection{Relativistic plane-wave impulse approximation}
\label{RPWIA}
\begin{figure}[ht]
\includegraphics[width=9cm]{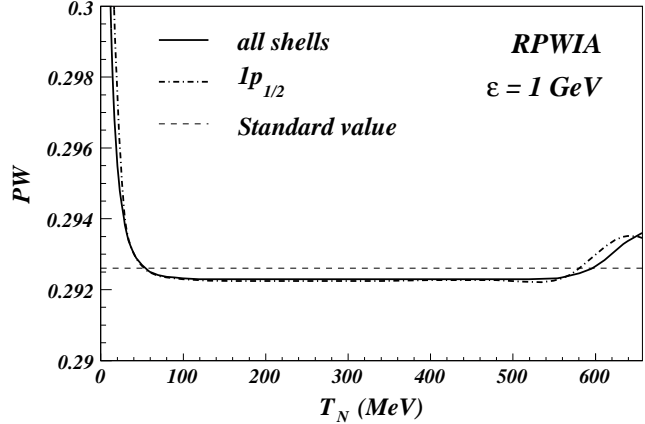}
\caption{The RPWIA Paschos-Wolfenstein relation as a function of the
  outgoing nucleon's kinetic energy $T_N$ for an incoming neutrino
  energy of $1$ GeV and an $^{16}$O target nucleus (full line).  Shown
  separately is the contribution of the $1p_{1/2}$ shell (dash-dotted).  The dashed line represents the analytic value 
derived in Eq.~(\ref{vier6}), with $\sin^2 \theta_W = 0.2224$ and $\cos \theta_c = 0.974$.}
\label{PaWo_RPWIA_1GeV}
\end{figure}

\begin{figure}[ht]
\includegraphics[width=9cm]{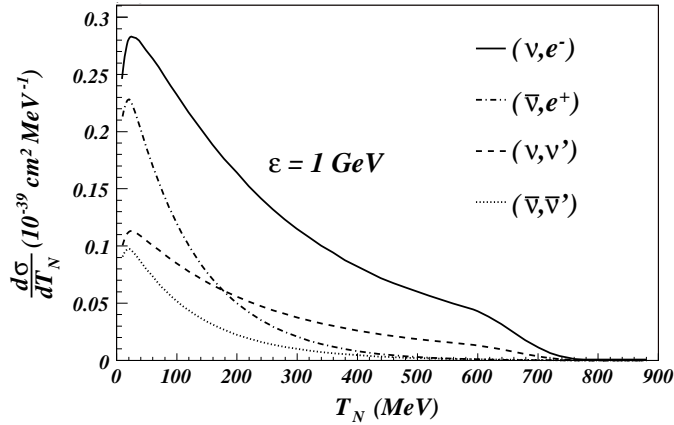}
\caption{$^{16}\mbox{O}$ differential cross sections for an incoming (anti)neutrino energy of $1$ GeV.  The full (dash-dotted) line represents the (anti)neutrino CC cross section, while the dashed (dotted) line depicts the (anti)neutrino NC cross section.}
\label{Ingredients_1000}
\end{figure}

\begin{figure}[ht]
\includegraphics[width=9cm]{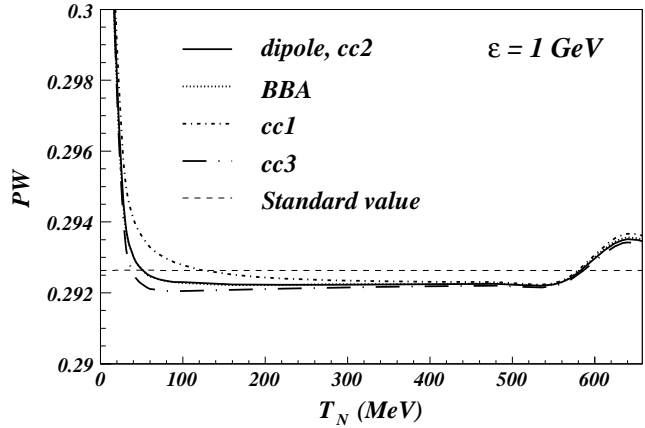}
\caption{The RPWIA Paschos-Wolfenstein relation as a function of $T_N$
  for the $^{16}$O $1p_{1/2}$ shell and an incoming neutrino energy of
  $1$ GeV.  The reference curve, with dipole vector form factors and the
  \textit{cc2} prescription for the one-nucleon vertex function, is drawn as a full line.
  Using the BBA-2003 parameterization results in the dotted curve.
  The (long) dash-dotted curve is obtained with the (\textit{cc3})
  \textit{cc1} prescription.  The dashed line represents the analytic value of Eq.~(\ref{vier6}).}
\label{PaWo_BBA_CCs}
\end{figure}

\begin{figure}[h]
\includegraphics[width=9cm]{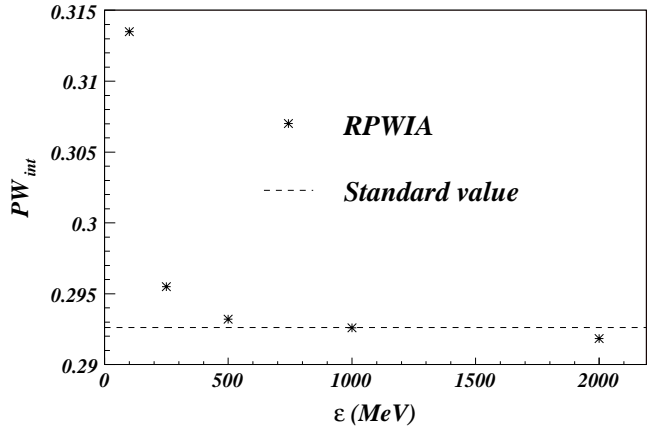}
\caption{Paschos-Wolfenstein relation for total $\nu/\overline{\nu}-^{16}$O cross sections against incoming neutrino energy.  The dashed line represents the standard value.}
\label{Integrated}
\end{figure}

Ignoring FSI of the ejectile with the residual nucleus, we adopt the
relativistic plane-wave impulse approximation (RPWIA).  Figure
\ref{PaWo_RPWIA_1GeV} displays the PW relation against the outgoing
nucleon's kinetic energy $T_N$ for an incoming neutrino energy of $1$
GeV and an $^{16}_{\phantom{0}8}\mbox{O}$ target nucleus.
Clearly, the $1p_{1/2}$-shell contribution to the PW relation can not
be distinguished from the total, shell-summed expression.  Both curves
show a remarkably constant behavior over a broad $T_N$ interval and
are in excellent agreement with the analytic value in
Eq.~(\ref{vier6}).  At very small $T_N$ values, threshold effects
induce large deviations.  The sudden increase near $T_N
\approx 550$\ MeV relates to a decrease of the corresponding
neutrino-induced NC and CC cross sections at the same energy, as shown
in Fig.~\ref{Ingredients_1000}.  For an incoming neutrino energy of $1$ GeV, 
nuclear binding effects do not seem to influence the PW relation
considerably.  As can be appreciated from Fig.~\ref{PaWo_RPWIA_1GeV},
Eq.~(\ref{vier6}) provides a very good approximation under those
circumstances.  In Fig.~\ref{PaWo_BBA_CCs}, we studied the sensitivity to the adopted
parameterization for the electroweak form factors. 
Employing the \textit{updated} BBA-2003 parameterization \cite{Budd1} for the weak
vector form factors apparently yields no difference with respect to the usual
dipole form.  Indeed, the fact that the results in
Figs.~\ref{PaWo_RPWIA_1GeV} and \ref{PaWo_BBA_CCs} are relatively $T_N$
independent indicates that the $Q^2$ dependence is largely cancelled
out in the PW ratio.  Accordingly, the sensitivity to the adopted
$Q^2$ evolution of the form factors is minor.  An interesting by-product of this feature is that the PW
relation does not depend on the axial form factor's cut-off mass
$M_A$, which constitutes a possible source of uncertainty in the
determination of $g_A^s$ from neutrino cross-section ratios
\cite{Alberico2, Alberico3}.  Similarly, Fig.~\ref{PaWo_BBA_CCs} shows
that the use of a different prescription for the weak one-nucleon current
operator exercises only the smallest of influences on the PW relation.\\ 
Most neutrino experiments, however, do not possess the discriminative
power to measure the ejectile's kinematics.  A comparison with
experimental results is facilitated using total cross sections, summed
over all final states of the outgoing nucleon.  Hence, it is useful 
to evaluate the integrated expression
\begin{equation}
\label{res2}
\mbox{PW}_{int} = \frac{\sigma^{NC}(\nu A) - \sigma^{NC}(\overline{\nu} A)}{\sigma^{CC}(\nu A) - \sigma^{CC}(\overline{\nu} A)},
\end{equation}                                      
obtained by integrating $d\sigma/dT_N$ over $T_N$.  Figure
\ref{Integrated} displays $\mbox{PW}_{int}$ for $\nu/\overline{\nu}\mbox{-}^{16}\mbox{O}$ 
cross sections and various incoming neutrino energies ranging from 
$100$ MeV to $2$ GeV. 
From $\epsilon = 500$ MeV onwards, the calculated values agree with
the standard value at the $0.5$ percent level, illustrating once more
the validity of the approximation of Eq.~(\ref{vier6}) in the
relativistic plane-wave approximation.  However, large discrepancies
are observed at lower incoming energies.  There, binding effects 
play an important role in the relative magnitude of the individual 
shell contributions to the cross sections.  As a result, the
expressions in numerator and denominator of Eq.~(\ref{vier3}) do not
cancel entirely, thereby shifting $\mbox{PW}_{int}$ to larger values.  With
increasing incoming neutrino energies, differences between the
contributions of different shells become of less importance and the
numerically computed PW values take on the value for the free nucleon.\\ 
In several experiments, $\nu_{\mu}$ and $\overline{\nu}_{\mu}$ beams
are employed.  Consequently, the outgoing-muon's mass needs 
to be taken into account when calculating the CC cross sections.
\begin{figure}[h]
\includegraphics[width=9cm]{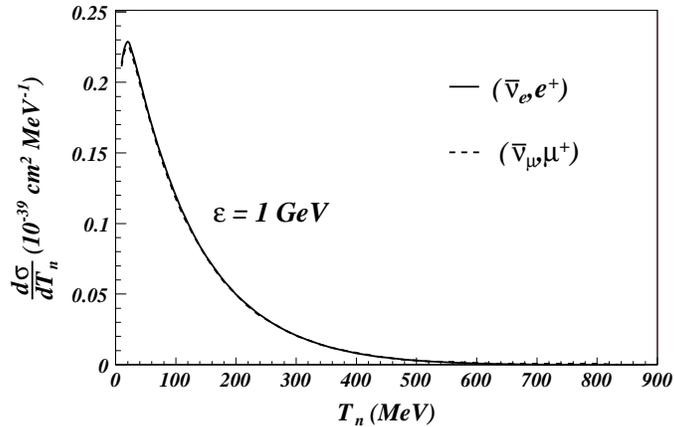}
\caption{Antineutrino-induced CC differential cross sections for
  $^{16}\mathrm{O}$ as a function of the outgoing neutron's kinetic
  energy $T_n$.  The full (dashed) line corresponds to an outgoing
  positron (antimuon).}
\label{cc-muon}
\end{figure}
For sufficiently high muon-neutrino energies, however, it is
readily seen that the mass of the muon ($\approx 105.7\ \mbox{MeV}$)
hardly influences the $T_N$ dependence of the CC cross sections.
Indeed, the nuclear responses should not be different, since a final
nucleon state of fixed kinetic energy must be created, irrespective of
the outgoing lepton's nature.  As for the kinematic factors (Table~I),
to a very good approximation the expression $\zeta = \sqrt{1 -
  \frac{M_l^2}{\epsilon'^2}}$ equals $1$ for electrons.  For
sufficiently high incoming energies, $\zeta \approx 1$ also holds for
muon neutrinos.  Figure \ref{cc-muon} indicates that this reasoning is
already valid for an incoming $\overline{\nu}_{\mu}$ energy of $1$ GeV.  
\subsection{Final-state interactions}
Unavoidably connected with the nucleon knockout channel under
consideration, is the nuclear effect stemming from the ejectile 
searching its way through the residual nucleus.  Here, these
final-state interactions are modelled by a relativistic
multiple-scattering Glauber approximation (RMSGA), introduced in Section \ref{cs}.
\begin{figure}[h]
\includegraphics[width=9cm]{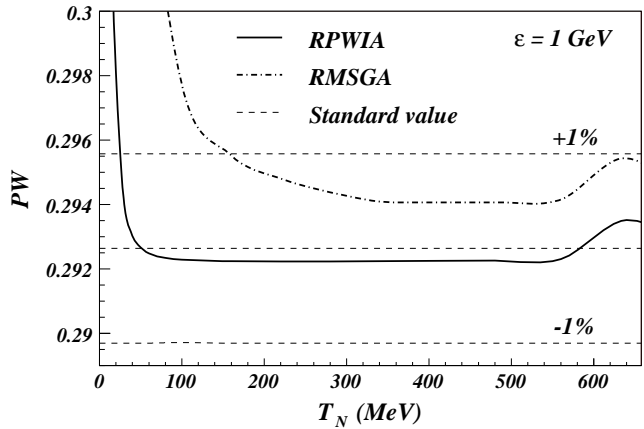}
\caption{The Paschos-Wolfenstein relation as a function of $T_N$ for
  the $^{16}\mbox{O}$ $1p_{1/2}$ shell.  The full (dash-dotted) line
  shows the RPWIA (RMSGA) case.  The dashed lines represent the
  standard PW value, with errors of $1\%$.}
\label{Glauber-1GeV}
\end{figure}   
In this Glauber model, FSI roughly halve the cross sections for
$^{16}\mbox{O}$.  As the PW relation takes ratios
of cross sections, FSI effects cancel to a large extent, which is
shown in Fig.~\ref{Glauber-1GeV} for an incoming neutrino energy
of $1$ GeV.  To better illustrate the influence of FSI mechanisms, a
$\pm 1 \%$ error on the standard PW value is indicated.  In
the region where the RMSGA produces valid results, i.e. for $T_N\
\mbox{down}\ \mbox{to}\ 200$ MeV \cite{Lava1}, FSI mechanisms increase 
the computed PW ratio by less than one percent.             
\subsection{Neutron excess}
\label{NE}
\begin{figure}[ht]
\includegraphics[width=9cm]{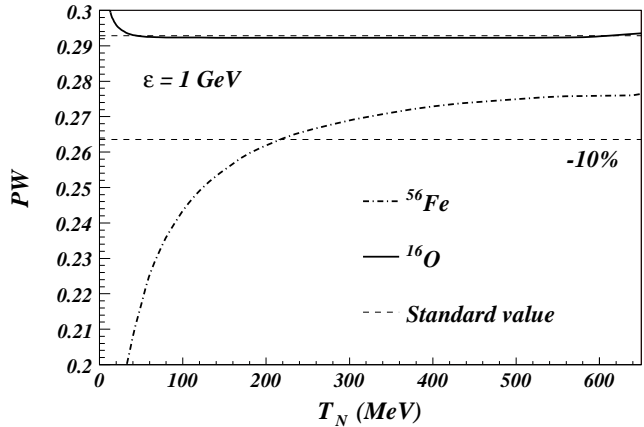}
\caption{The RPWIA Paschos-Wolfenstein relation as a function of $T_N$
  for an iron target (dash-dotted).  Other notations refer to
  Fig. \ref{PaWo_RPWIA_1GeV}.  For reference purposes, a dashed-line 
denoting the $10\%$-reduced standard PW value is added.}
\label{Iron-1GeV}
\end{figure}

In the preceding subsections, the PW relation was investigated for a
target with an equal number of protons and neutrons.  For sufficiently high energies, the balance between protons
and neutrons make the $\sin^2 \theta_W$ dependence of the PW relation
the traditional one of Eq.~(\ref{pw5}).  Evidently,
neutrino-scattering experiments often employ heavier target
nuclei, with an excess amount of neutrons.  The additional
energy-dependent terms that are introduced in the PW formula will
affect the predicted PW standard value (\ref{vier6}), which required 
the perfect cancellation between proton and neutron contributions.  
Figure \ref{Iron-1GeV} shows the $T_N$ dependence of the PW relation for
$^{56}$Fe at an incoming neutrino energy of $1$ GeV.   
The specific energy dependence of PW in the iron case is given shape
by the extra $\nu$-induced CC cross sections in the denominator.  Thereby, low
PW values correspond with the peak region and high values with
the tail of the excess neutrons' contribution to $\sigma^{CC}(\nu A)$.
In general, the neutron excess in the iron target lowers PW values 
by $\gtrsim 10\%$.  Correspondingly, of all nuclear
effects looked into here, the neutron-excess correction to the PW
relation is the largest and most important one.     
\subsection{Model dependence and $\sin^2 \theta_W$ determination}
Of course, to be relevant for future neutrino-scattering experiments,
the above predictions need to be discussed in terms of their model
dependence.  To this end, we follow the line of reasoning in
Refs.~\cite{Alberico1, Alberico3}, where the difference between 
cross sections provided by a relativistic Fermi-gas model (RFG)
and a relativistic shell model (RSM) is assumed to be indicative of
the theoretical model uncertainty itself.  While sizeable for separate
cross sections at lower incoming neutrino energies, nuclear-model
dependences already seem to vanish at $\epsilon = 1$ GeV where the RSM
curves coincide with the RFG ones \cite{Alberico1}.  
A similar conclusion is reached in \cite{Lava1}, where a comparison is
made between RPWIA shell-model cross sections and RFG results.  As the neutrino energy increases to $1$ GeV,
the RFG curves approach more and more the RPWIA predictions.  In the same
work, two methods to incorporate FSI mechanisms were compared: the Glauber
approach also applied here and the relativistic optical potential approximation.  At $\epsilon = 1$
GeV, both techniques were found to produce similar results down to
remarkably low nucleon kinetic energies $T_N \sim 200$\ MeV.  Hence, 
as nuclear-model uncertainties seem to be negligible at $\epsilon = 1$
GeV for separate cross sections, we conclude that the PW relation, a
superratio, mitigates these model dependences well below the level of all
other nuclear effects studied in this work.\\
For isoscalar target nuclei and energetic neutrinos, 
the whole of nuclear-model uncertainties on the PW relation is 
seen to be well within percentage range.  Evidently, this means that a
PW measurement with percent-level accuracy can only resolve
non-isoscalar nuclear effects.  Notwithstanding the extreme stability 
with respect to theoretical uncertainties in nuclear modelling, a quick glance at the
PW relation's Weinberg-angle sensitivity (from Eq.~(\ref{pw5})) 
\begin{equation}
\label{sens}
\frac{\Delta \mbox{PW}}{\mbox{PW}} = \frac{- \Delta \sin^2 \theta_W}{\frac{1}{2} -
  \sin^2 \theta_W},
\end{equation}     
immediately qualifies any ambition to exploit the PW relation as an electroweak
precision tool.  From Eq.~(\ref{sens}), a $\pm1\%$ theoretical
uncertainty on the PW relation would result in an equally large
\textit{nuclear-model error} on the Weinberg angle
$\Delta_{nuc}(\sin^2 \theta_W) = \mp 0.0028$.  On
the contrary, a 10\% measurement error for the parity-violating asymmetry $A_{PV}$
in $\vec{e}e$ M{\o}ller scattering at $Q^2 = 0.026\ \mbox{GeV}^2$
translates in a $1\%$ uncertainty on the corresponding Weinberg-angle
value \cite{Anthony1}.  The newly
proposed Qweak experiment at Jefferson Lab aims at a $4\%$ measurement
of the proton's weak charge $Q^{p}_{w}$, resulting in a $0.3\%$
measurement of $\sin^2 \theta_W$ \cite{Qweak1}.  In this type of
experiments, the sensitivity to the weak mixing angle is substantially
enhanced by the factor $1/4 - \sin^2 \theta_W$ figuring in the
$A_{PV}$ expression.  Obviously, the PW relation cannot compete with
the level of sensitivity achievable in this sector and is therefore
less suited as an electroweak precision test.              

\subsection{Strangeness}
\label{Strange}
\begin{figure*}[t]
\includegraphics[width=18cm]{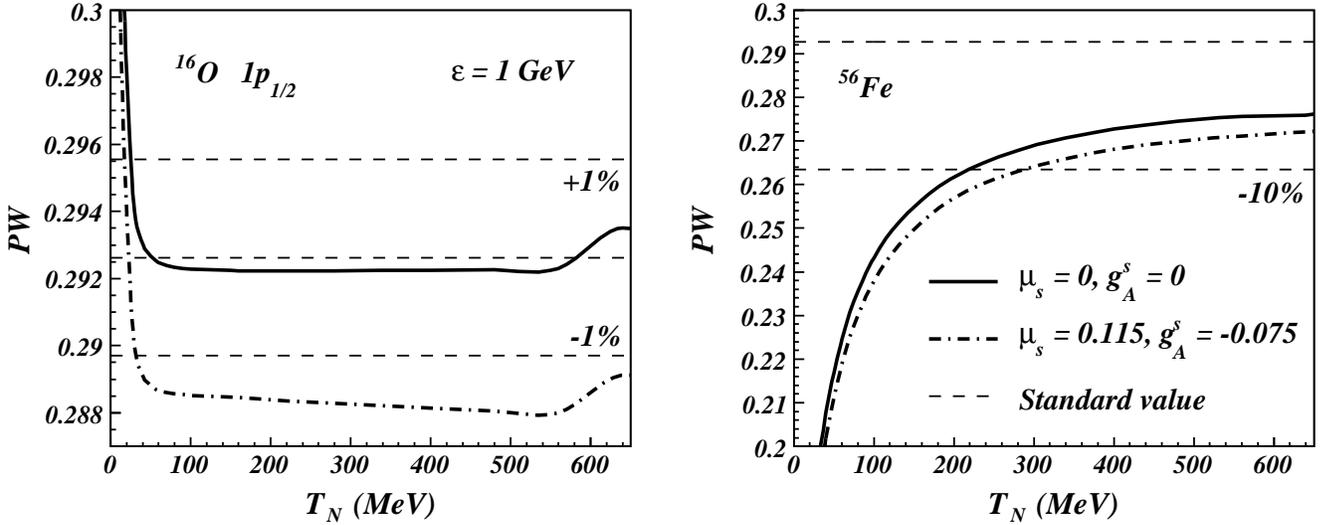}
\caption{The left (right) panel shows the RPWIA Paschos-Wolfenstein
  relation for the $^{16}\mbox{O}$ $1p_{1/2}$ shell (an
  $^{56}\mbox{Fe}$ target nucleus) and a $1$ GeV incoming neutrino
  energy.  Full (dash-dotted) lines correspond to $g_A^s = \mu_s = 0$
  ($g_A^s = -0.075, \mu_s = 0.115$).  For comparison, the standard PW
  values without strangeness are included (dashed lines).}
\label{Strange1}
\end{figure*}

\begin{figure}[h]
\includegraphics[width=18cm]{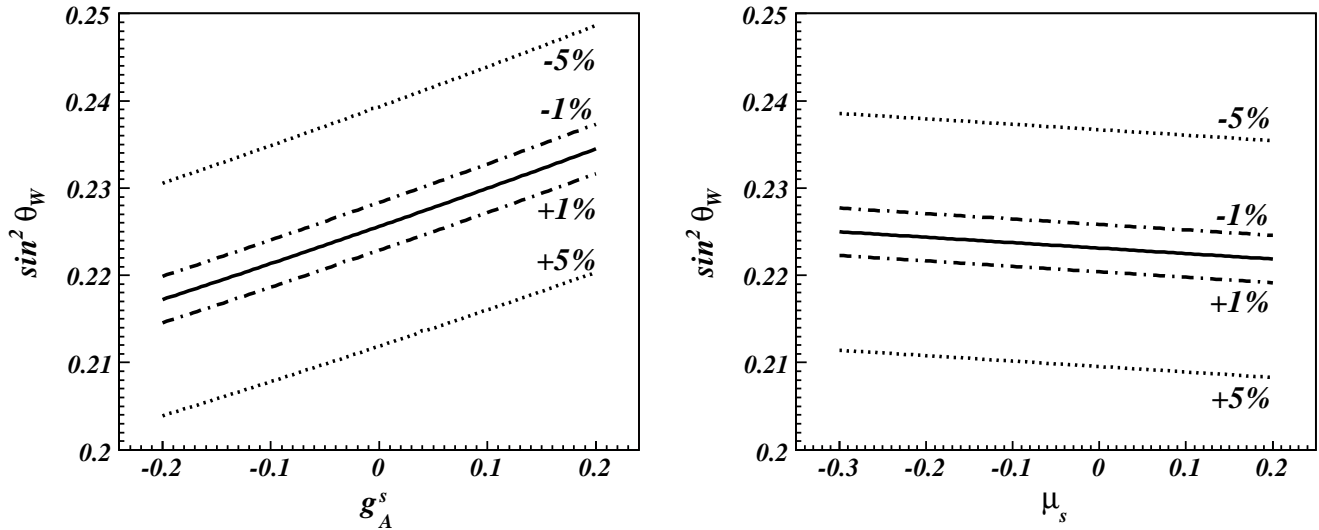}
\caption{Plots showing how $\sin^2 \theta_W$ and strangeness parameter
  values are correlated in the PW relation.  The full line corresponds 
to values of the indicated parameters for which the PW relation is
  constant.  The dash-dotted (dotted) lines have the same meaning, but
with PW equal to $\pm 1\%$ ($\pm 5\%$) the full-line value.}
\label{Strange2}
\end{figure}

\begin{figure}[h]
\includegraphics[width=18cm]{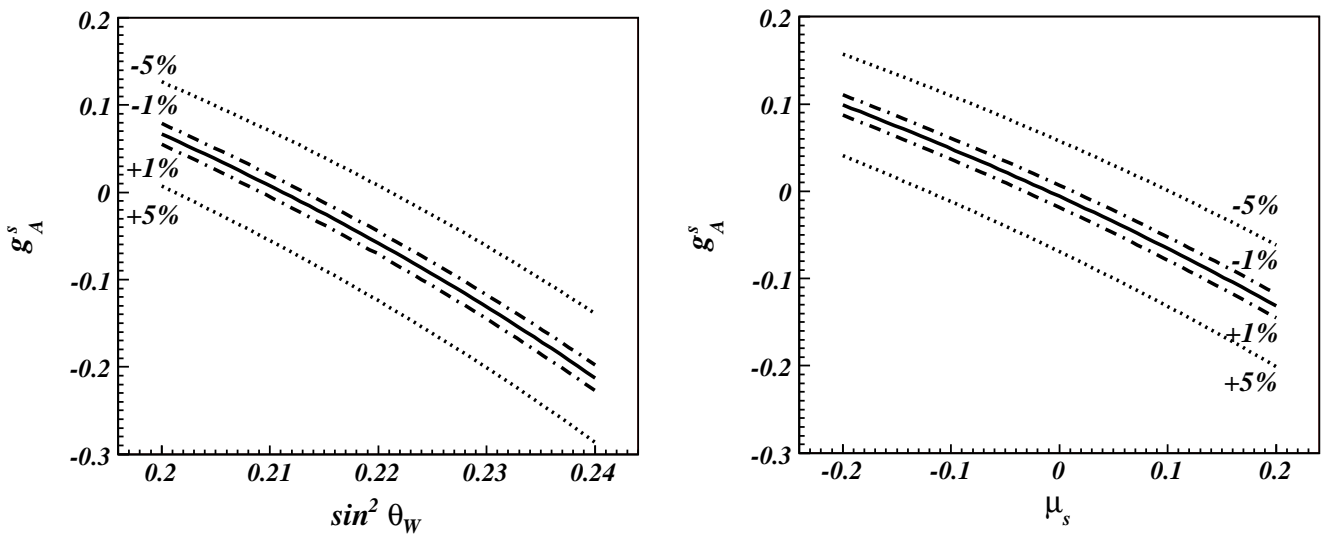}
\caption{Correlation plots showing how the axial strangeness parameter
  $g_A^s$ is intertwined with $\sin^2 \theta_W$ (left) and $\mu_s$
  (right) through the PW relation for proton knockout $\mbox{PW}_{p}$.  The full line corresponds 
to values of the indicated parameters for which $\mbox{PW}_{p}$ is constant.  The dash-dotted (dotted) lines have the same meaning, but
with $\mbox{PW}_{p}$ equal to $\pm 1\%$ ($\pm 5\%$) the full-line value.}
\label{Strange3}
\end{figure}

As a final point, we discuss the impact of the nucleon's strangeness
content on the PW relation.  State-of-the-art reviews addressing 
the experimental progress on strange electromagnetic form factors and
the strangeness contribution to the nucleon's spin can be found in 
Refs.~\cite{MusolfStr} and \cite{PateStr} respectively.  Generally
speaking, PVES experiments show a tendency towards small, positive
values for the strangeness magnetic moment $\mu_s$ \cite{MusolfStr,
  G0, HAPPEX}.  
Leptonic DIS experiments seem to suggest a value of $\approx -0.1$ 
for $g_A^s$ \cite{PateStr}.  As baseline strangeness parameter values, 
we therefore adopt predictions from the chiral quark-soliton model (CQSM) 
with kaon asymptotics \cite{Silva1}, namely $\mu_s = 0.115$ and $g_A^s =
-0.075$.  We wish to stress that the available strangeness
information still exhibits relatively large error flags.
Moreover, there exist fundamental discrepancies between the
experimentally favored positive $\mu_s$ and most model predictions
\cite{MusolfStr, Lava2}.  So, the values used here can be regarded 
as a model prediction for $\mu_s$ and $g_A^s$ which is compatible with
currently available data.  Figure \ref{Strange1} illustrates the
influence of non-zero strangeness parameters on the PW relation.                 
As can be observed from the left panel, the inclusion of strangeness 
alters the PW relation for an isoscalar target by an amount of $\sim
1\%$.  For $^{56}\mbox{Fe}$, a nucleus with neutron excess, the effect 
is larger ($\sim 2\%$).  Summing over an equal number of proton and
neutron contributions effectively cancels all isovector-strangeness
interference terms, thereby reducing the PW relation to the analytic
estimate (\ref{vier6}).  On the contrary, the extra neutrons in
$^{56}\mbox{Fe}$ skew this proton-neutron balance, producing a larger 
deviation from the PW relation without strangeness.  Clearly,
strangeness adds a significant amount of uncertainty when attempting
to determine $\sin^2 \theta_W$ from the PW relation.  A simple way of
visualizing the mutual influence of the parameters entering 
the PW relation is by considering the correlation plots in
Fig.~\ref{Strange2}.  We took Eq.~(\ref{vier6}) with the baseline
parameter values as a starting point to calculate the lines of
constant PW.  From the left panel of Fig.~\ref{Strange2}, one can infer 
that a $50\%$ uncertainty on $g_A^s$ translates in a $0.7\%$ error 
on $\sin^2 \theta_W$ if we assume that everything else is known.  On the other hand, 
extracting $\sin^2 \theta_W$ from the PW relation is visibly 
less sensitive to the value of $\mu_s$, yielding only a $+0.3\%$
increase if $\mu_s$ is changed from $0.115$ to $0$.   
Again, it emerges that the limited information on $g_A^s$ and $\mu_s$
presently at hand, does not allow one to exploit the PW relation to
probe the Weinberg angle with the sensitivity achievable in PVES.  
Turning things around, however, a precisely known Weinberg-angle value may turn
out valuable in trying to pin down $g_A^s$ from a measurement of the
QE PW relation.  Ratios of neutrino-induced cross sections are indeed 
considered useful for studying the strangeness content of the nucleon,
and notably the strangeness contribution to the nucleon's spin
$g_A^s$.  Well-covered examples are the ratio of proton-to-neutron NC reactions
\cite{Garvey1, Alberico1, Alberico3}, NC to CC cross-section ratios \cite{vanderVentel1,
  Meucci1}, polarization asymmetries \cite{Lava2} and the
Paschos-Wolfenstein relation for proton knockout $\mbox{PW}_{p}$ \cite{Alberico1}.  In the
latter article, $\mbox{PW}_{p}$ was seen to have a strong dependence on
$g_A^s$.  In addition, results presented in this work justify the
optimism about a model-independent $g_A^s$ determination \cite{Alberico2} by measuring
$\mbox{PW}_{p}$ in the right circumstances, i.e. with an isoscalar target
nucleus and an incoming neutrino energy of about $1$ GeV.  To study how
the finite precision on $\sin^2 \theta_W$ and $\mu_s$ influences the
accuracy with which $g_A^s$ can be extracted from $\mbox{PW}_{p}$, we
consider the correlation plots in Fig.~\ref{Strange3}.  The curves
were again drawn from Eq.~(\ref{vier6}), now retaining only the proton
contribution in the numerator ($\tau_3 = +1$) to obtain lines of
constant $\mbox{PW}_{p}$.  From this figure, we see that a $5\%$ measurement of
$\mbox{PW}_{p}$ results in a $\pm 0.067$ determination of $g_A^s$.  For
comparison, the FINeSSE collaboration aims at a $6\%$ measurement of the
NC/CC ratio down to $Q^2 = 0.2\ \mbox{GeV}^2$, corresponding to a $\pm
0.04$ measurement of $g_A^s$.  The left panel in Fig.~\ref{Strange3}
learns that a $1\%$ uncertainty on $\sin^2 \theta_W$ gives rise to 
a $20\%$ uncertainty on $g_A^s$, assuming again that everything
else is fixed.  The inconclusive information on 
$\mu_s$ available at present has a far more severe effect on the value
of $g_A^s$, as can be derived from the right panel.  Shifting the 
strangeness magnetic moment from $0.115$ to $0$, $g_A^s$ changes by
$\sim 0.07$.  We recall that nuclear-model uncertainties can be
mitigated to the $1\%$ level, corresponding to $\Delta_{nuc}(g_A^s)
\sim 0.015$.  This analysis stresses the importance of further
experimental efforts to put more stringent limits on the 
strangeness form factors of the nucleon.  As apparent from this
$\mbox{PW}_{p}$ case, experiments in the vector and axial-vector sector 
heavily depend on each other in the sense that both types of
measurements need reliable input values for the other strangeness
parameters.                                       

\section{Conclusions}
Adopting a fully relativistic nucleon knockout model for the
description of quasi-elastic neutrino-nucleus interactions, we have
conducted a study of the Paschos-Wolfenstein relation with hadronic
degrees of freedom.  Results are presented for $^{16}$O and
$^{56}$Fe target nuclei and incoming neutrino energies between $100$
MeV and $2$ GeV.  We estimate that nuclear-model uncertainties 
amount to a $1\%$ theoretical error bar for the PW relation in the case of
sufficiently high neutrino energies ($\gtrsim 1$\ GeV) and
isoscalar target nuclei.  Under these conditions, the Weinberg-angle
dependence of the PW relation is to a very good approximation
identical to the one constructed with DIS neutrino-nucleon cross sections.  
Binding effects produce a sizeable shift at lower incoming neutrino
energies, but become negligible beyond $500$ MeV.  The largest correction
stems from neutron excess in the target, which drastically lowers the
PW curve.  Though nuclear-model effects are extremely well controlled,
the PW relation is no match for electroweak precision probes in other
sectors, notably PVES experiments whose sensitivity to the Weinberg
angle is considerably larger.  The poor information on the nucleon's
strangeness content presently at hand also induces $1\%$-level
uncertainties on the PW relation, and consequently puts even more
stringent limits on its $\sin^2 \theta_W$ sensitivity.  An extraction 
of the strangeness contribution to the nucleon's spin $g_A^s$ through 
the proton knockout part of the PW relation clearly benefits 
from the small theoretical uncertainties involved ($\Delta_{nuc}(g_A^s)
\sim 0.015$), but depends heavily on a reliable input for the strange
vector form factors.
                     
\acknowledgments
The authors acknowledge financial support from the Fund for Scientific Research (FWO) Flanders and the University Research Board (BOF).

\end{document}